\documentstyle[12pt,aaspp4]{article}
\def\Ha{H$\alpha$}
\def\Hb{H$\beta$}
\def\MgIUV{Mg{\sc\ I} $\lambda$2852}
\def\MgII{Mg{\sc\ II} $\lambda\lambda$2795,2802}
\def\OID{[O{\sc\ ~I}] $\lambda\lambda$6300,6364}
\def\OIDB{[O{\sc\ I}] $\lambda$6300}
\def\OIDR{[O{\sc\ I}] $\lambda$6364}
\def\OIIUV{[O{\sc\ II}] $\lambda\lambda$3726, 3728}
\def\OIIIR{[O{\sc\ II}] $\lambda\lambda$7319, 7330}
\def\MgIO{Mg{\sc\ I}] $\lambda$4571}
\def\MgIUV{Mg{\sc\ I} $\lambda$2852}
\def\MgIIUV{Mg{\sc\ II} $\lambda\lambda$2795,2802} 
\def\ROID{R\,=\,F([{\rm OI}]6300)/F([{\rm OI}]6364)}
\def\Msun{${\rm M_\odot}$}
\def\kms{$\rm{\,km\,s^{-1}}$}
\def\arcsec{$^{\prime\prime}$}
\begin{document}

\title{
  Hubble Space Telescope Spectroscopic Observations of the Ejecta of SN
  1987A at 2000 Days\footnote{Based
  on observations with the NASA/ESA Hubble Space Telescope, obtained at the
  Space Telescope Science Institute, which is operated by AURA, inc., under 
  NASA contract NAS 5-26555}}

\author{Lifan Wang\footnote{Also Beijing Astronomical Observatory, 
Beijing 100080, P. R. China}, 
        J. Craig Wheeler}
\affil{Dept. of Astronomy and McDonald Observatory,
          The University of Texas at Austin,\\
          Austin, TX~78712}
\authoremail{lifan@astro.as.utexas.edu, wheel@astro.as.utexas.edu}
\author{Robert P. Kirshner, Peter M. Challis}
\affil{Harvard-Smithsonian Center for Astrophysics, 60 Garden Street, MS-19, 
       Cambridge, MA 02138}
\author{Alexei V. Filippenko}
\affil{Department of Astronomy, University of California, Berkeley, CA~94720}
\author{Claes Fransson}
\affil{Stockholm Observatory, S-133 00 Saltsj\"obaden, Sweden}
\author{Nino Panagia\footnote{Affiliated to the Astrophysics Division, Space
        Science Department of ESA.}}
\affil{Space Telescope Science Institute, 3700 San Martin Drive, Baltimore, 
       MD~21218}
\begin{center}
and
\end{center}
\author{Mark M. Phillips, Nicholas Suntzeff}
\affil{Cerro Tololo Inter-American Observatory, National Optical Astronomy
       Observatories, Casilla~603, La Serena, Chile}
\begin{abstract} 

We have used the Faint Object Spectrograph on the Hubble Space Telescope
to observe the spectra of SN 1987A over the wavelength range 
2000 -- 8000\ \AA\ on dates 1862 and 2210 days after the supernova outburst. 
Even these pre-COSTAR
observations avoid much of the contamination from the bright stars nearby and
provide a very useful set of line strengths and shapes for analysis. The 
spectrum is formed in an unusual physical setting: cold gas which is excited
and ionized by energetic electrons from the radioactive debris of the supernova
explosion. The spectra of SN 1987A at this phase are surprisingly 
similar to those of the nova shells of CP Puppis and T Pyxidis decades 
after outburst. SN 1987A and the novae are characterized by emission from 
material with electron temperatures of only a few hundred degrees Kelvin, 
and show narrow Balmer continuum emission and strong emission lines from O$^+$. 
The Balmer continuum shape requires the electron temperature in the 
supernova ejecta to be as low as 500 K on day 1862 and 400 K on day 2210 
after outburst. The \OIIUV\ doublet is surprisingly strong and is plausibly
powered by collisional ionization of neutral oxygen to excited states of 
O$^+$. 

The line intensity ratio of the \OID\ doublet obtained from Gaussian 
fits of the line profiles is 1.8$\pm0.2$, contrary to the optically thin 
limit of 3. This ratio is {\it not} due to an optical depth effect, but 
rather is an artifact of assuming a Gaussian profile to fit the \OID\  
doublet profile. Specifying the line ratio 
$R\, = \, F([{\rm OI}]6300)/F([{\rm OI}]6364)$ = 3 is consistent with the
data and allows a calculation of the decomposed line profile. 

All the observed strong lines  are found to be blueshifted by a similar 
amount of 400 \kms. The line profiles are quite similar for lines arising 
from different chemical elements. The profiles are all asymmetric, showing 
redshifted extended tails with velocities up to 10,000 \kms\ in some strong 
lines. The blueshift of the line peaks is attributed to dust that 
condensed from the SN 1987A ejecta which is still distributed in dense 
opaque clumps. The strongest ultraviolet lines are those of \MgIUV\ and 
\MgIIUV. The \MgIUV\ line is significantly broader than most lines in the 
optical, which provides a natural explanation for the size differences in 
the optical and ultraviolet of the SN 1987A ejecta derived from HST direct 
images. 
\end{abstract}


\section{Introduction}

SN 1987A provides a unique opportunity to observe the debris from a supernova
explosion. As part of our Supernova INtensive Study (SINS) of supernovae with 
HST,
we present spectroscopic observations obtained with the Hubble Space Telescope
(HST). When these HST observations were taken, in 1992 and 1993, the ejecta 
were so cold that thermal electrons played a negligible role in the formation 
of optical and ultraviolet lines. The observed emission lines are powered 
through non--thermal excitations by fast electrons produced by collisions 
with the gamma rays that result from the decay of radioactive $^{57}$Co and 
$^{44}$Ti produced in the supernova explosion. Detailed models for 
the spectroscopic evolution of supernova ejecta in the nebular phase have 
been presented by several groups (Fransson \& Chevalier 1989; Swartz 1991; 
Xu \& McCray 1991;  Kozma \& Fransson 1992; Li \& McCray 1995). An updated
review paper is given by McCray (1993). The basic mechanism 
involves energy deposition by gamma rays from radioactive materials 
($^{56,57}$Co and $^{44}$Ti) in the supernova ejecta. The fast electrons 
produce heating, excitation, and ionization. Five years after the 
supernova explosion, excitation and ionization dominate the energetics, 
while heating becomes unimportant due to the low ionization fraction of 
the ejecta. For most ions, 
recombination time scales in the SN 1987A ejecta at late times may exceed 
the dynamical and radioactive decay time scale and the ionization stage of 
the ejecta will freeze out (Clayton, Leising, \& The 1992; Fransson \& 
Kozma 1993). As a result, emission lines from radiative recombination 
remain strong long after the supernova outburst. Late time 
observations of SN 1987A provide us with an excellent opportunity to 
test these theories, and to analyze the ejecta.

The purpose of this paper is to present the pre-COSTAR HST observations of 
SN 1987A, present plausible explanations of the physical processes, and to 
derive some properties of the underlying event. We discuss in \S 2 the 
observations and data reduction. The results from this study are given in 
\S 3, where we discuss emission lines arising from hydrogen, oxygen, magnesium
and iron, and the properties of the dust in SN 1987A. We give  
conclusions from this study in \S 4. 

\noindent
\section{Observations and Data Reduction}

The spectra of SN 1987A were acquired using the HST as part of the Supernova 
INtensive Study. Observations with the FOS were obtained on 1992 March 31, 
1993 March 14, and 1993 April 1-6.  Post-COSTAR observations were made on 
1995 Jan. 7, and will be the subject of a subsequent paper. In addition to 
these spectroscopic observations, there are also WFPC images taken as part 
of the same project. The direct images have been discussed by Plait et al. 
(1994) and by Pun et al. (1996). This paper concentrates on the two pre-COSTAR spectra. 

The data were reduced using the standard HST post-processing software. 
The pipeline-calibrated data were reduced with the most current 
calibration reference files. Table 1 contains the summary of the 
observations including dataset name, date of observation, detector, 
grating, and exposure time. Table 2 contains the observational parameters 
including wavelength coverage, resolution, and dispersion. All the data 
were taken through the 0\arcsec.25 $\times$ 2\arcsec.0 slit. The slit was 
located to observe 
both the broad lines from the debris and narrow lines from the circumstellar 
ring simultaneously. Figure 1 shows the slit orientation and position 
for each set of observations. Both the 1992 March 31 and 1993 March 14 
observations were centered on the debris. Figure 1 was created from a FOC 
observation of SN 1987A on 1992 April 12.  The 1993 April 1-6 observations 
were not centered on the target, our best estimate is that the slit was 
positioned $\sim$0\arcsec.3 east of the debris. These miscentered 
observations show spectral features of the circumstellar ring, but no 
features from the debris. A new acquisition strategy has since been 
implemented to insure the aperture is well centered on SN 1987A.

At the off axis location of the FOS, the width (FWZI) of the Point Spread 
Function (PSF) of the optical system is about 2-3\arcsec. Figure 1 shows 
that the light 
from star 3 (1\arcsec.3 away) overlaps the position of the FOS aperture. Since 
star 3 is a Be star (Wang et al.. 1992; Walborn et al., 1993), the 
contamination is most serious in the UV. The amount of star 3 light in the 
supernova spectra can be estimated by two methods. 
First, the FOC images can be used to determine how much of the star 3 light 
enters the FOS aperture. Second, the FOS observations from 1995 Jan., during 
the COSTAR era, can be used as a guide to adjust the pre-COSTAR data.
A correction can be calculated by measuring the difference in the flux
level between the emission lines of the pre-COSTAR and the  post-COSTAR 
spectra. Both methods give a consistent correction to the data. 
The contamination of the observed spectra by star 3 varies with wavelength,
and ranges from as large as 75\% at 2000\ \AA\  to 17\% at 4000\ \AA. At 
wavelengths longer than 5000\ \AA, this contamination is below 1\%.   
The observations made with the blue side of G130H FOS are completely 
dominated by Star 3 light. No emission features were detected from either 
the debris or the circumstellar ring.

\noindent
\section{Results}

The final flux calibrated spectra are corrected for interstellar extinction 
assuming $E(B-V)\,=\,0.2$ mag (Panagia et al. 1987) and $A_v\,=\,0.62$. The
spectra are shown in Figure 2 with the suggested line 
identifications. The original spectra consist of both narrow forbidden 
lines typical of HII regions and broad lines of FWHM 
greater than 2000 \kms. The narrow lines were first observed with the IUE 
(Fransson et al. 1989) in the ultraviolet about 80 days after the supernova 
explosion, and from the ground (Wampler \& Richichi 1989) in the optical 
about 300 days after the explosion. These lines are believed to be due to 
the circumstellar material in the immediate SN 1987A environment, presumably 
material lost by the progenitor's stellar winds. The broad lines are from 
the supernova debris itself, and are the main topic of the current study.
All the narrow lines have been removed from the spectra in Figure 2, but form
the subject of a future paper (Panagia et al. 1995).

A complete list of the emission lines from the debris is given in
Table 3. The wavelengths given in the table were approximate meassured values
of the line peaks. The line strengths were measured by fitting a 
simple Gaussian
profile with a linear background. There are three main difficulties in
measuring the line strengths. First, there are the narrow emission
features from the circumstellar ring blended with the broader debris
lines. These emission features were fit with a single Gaussian profile 
with a linear background and subtracted. Secondly, the debris emission 
features are not exactly Gaussian, as discussed in \S 4. Line strengths 
obtained by simply integrating above a continuum are accurate to within this 
uncertainty, and therefore only values from Gaussian fits are tabulated. 
Third, the continuum level is often difficult to measure as 
many of the lines are blended. The estimate of the uncertainty of the
line strength is 10--20 percent including also the uncertainties in
the calibration of the absolute flux level. 
However, the relative strengths of the lines are more reliable, and
are accurate to about 5 percent. The observed \Ha\ fluxes are 
1.66$\,\times\,10^{-13}\,{\rm erg\,cm^{-2}\,s^{-1}\,\AA^{-1}}$ and 1.17 
$\,\times\,10^{-13}\,{\rm erg\,cm^{-2}\,s^{-1}\,\AA^{-1}}$ for the 1992 and
1993 spectra, respectively. The dereddened \Ha\ fluxes for the 1992 and 1993 
spectra are 2.62$\,\times\,10^{-13}\,{\rm erg\,cm^{-2}\,s^{-1}\,\AA^{-1}}$
and 1.84$\,\times\,10^{-13}\,{\rm erg\,cm^{-2}\,s^{-1}\,\AA^{-1}}$, respectively.

Most of the lines which appeared in earlier nebular phase spectra, such as 
those identified by Phillips \& Williams (1991), and Kirshner et al. (1987)
are still detectable. The 
strongest lines are \Ha, \Hb, and \OID\ in the optical, and the resonance 
\MgIUV, \MgII\ lines discussed in detail by Pun et al. (1995) in the 
ultraviolet. A remarkable feature shown by the current data is that some 
of the iron lines, such as the [Fe II] $\lambda 7155$ line, which were strong 
in the early spectra have now disappeared. The temperature of the supernova 
ejecta is too low, and excitations by thermal electrons are no longer 
important for optical emission lines like [Fe II] $\lambda 7155$ at this 
epoch. As a consequence, most of the thermally excited emission lines 
should appear in the 
infrared where thermal excitations are still important. This behavior is 
related to the so-called infrared catastrophe as studied in detail by 
Fransson \& Chevalier (1989). 

An interesting problem is that there are still several strong lines without 
secure identification, notably the emission features at wavelength 3630\ \AA\, 
and 3727\ \AA. These lines will be discussed in more detail in the subsequent 
sections of this paper. We show that these are likely to be the Balmer 
continuum at low temperature and [O II] produced by ionization 
of neutral oxygen to excited states of O$^+$.

\noindent
\subsection{Hydrogen Emission}

\noindent
\subsubsection{\Ha\ and \Hb\ Lines}

Hydrogen Balmer lines remain the strongest optical lines. The profiles of 
\Ha\ and \Hb\ lines show an asymmetric redshifted tail out to 10000 \kms\ 
for \Ha\ and 7000 \kms\ for \Hb\ beyond which the line profile is 
corrupted by an iron line. The profiles are shown in Figure 3 for the two 
epochs of observation. Both the \Ha\ and \Hb\ profiles show the same 
asymmetry, so this is an intrinsic feature of the hydrogen, not a 
blending with other lines.

The asymmetric component in the line profile can be extracted by flipping 
the blue side of the \Ha\ profile about the line peak and subtracting it 
from the red wing of the line profile. This results in a bump as broad as 
8000 \kms\ (FWZI) centered at a redshift about 4000 \kms. The central velocity 
of this asymmetric component is in remarkable agreement with the analysis of 
the redshifted lines observed between day 200 and day 500 (Spyromilio et al.
1990), and the red emission satellite of the hydrogen lines between day 20 
and day 100, the asymmetry noted by Hanuschik \& Dachs (1987), Larsonet et 
al. (1987), Phillips \& Heathecote (1989) and Spyromilio et al. (1990).
Chugai (1991a, 1991b) suggested that this asymmetric component is due 
to an asymmetric distribution of radioactive clumps in the SN 1987A ejecta.

\noindent
\subsubsection{Hydrogen Balmer Continuum}

Another interesting feature is at wavelength 3630\ \AA. While this appears to 
be an emission line with width comparable to other lines, we identify it as 
hydrogen Balmer continuum emission, based on the argument developed by 
Williams (1982) in his spectroscopic analysis of the shells around 
novae. The nova shells Williams analyzed are quite cold, with electron 
temperature of only about 800 K. They emit unusual spectra in which recombination 
lines, including some forbidden lines formed by recombination, are quite 
strong. Because the emission coefficient for the Balmer continuum varies as 
$j_\nu\,\sim \, \exp[h(\nu_{\rm o}-\nu)/kT_e]$, where $\nu_{\rm o}$ is the Balmer 
ionization frequency, the Balmer continuum profile drops very sharply at 
frequencies higher than $\nu_{\rm o}$ when the kinetic temperature of the free 
electrons is very low, producing a narrow emission feature in the spectrum.

This identification of the Balmer continuum profile provides us with a 
powerful method for estimating the electron temperature in the ejecta. 
In order to do so, we have constructed a simple model which assumes that 
(a) there is a uniform distribution of hydrogen atoms inside a sphere of 
velocity smaller than 3000 \kms; (b) the temperature is constant across 
the entire nebula; (c) the ejecta are optically thin to Balmer continuum 
photons. Assumption (a) is required mainly because of the lack of 
knowledge of the density structure of the debris; however, it is perhaps not 
very far from reality considering the complex chemical mixing processes 
indicated by various observations and models (see, McCray, 1993 for a 
review of this topic). Assumption (b) implies that the temperature to be 
deduced is only an average value. Assumption (c) can be justified because 
the ejecta are now much more extended than in early phases of evolution, 
and are now transparent to Balmer continuum photons (e.g., 
McCray 1993).  

The blue side of the Balmer continuum profile is temperature sensitive, but 
is insensitive to the density structure. On the other hand, the 
redshifted portion of the Balmer continuum is determined primarily by the 
density structure of the ejecta, in which different velocities give 
different weights which alter the Doppler smearing of the Balmer edge. By 
matching both the blueshifted and redshifted parts of the profile, we are 
able to deduce the temperature of the ejecta quite accurately. The model 
profiles are plotted together with the observed line profiles in Figure 4. 
The corresponding best values of the electron temperature are 500 K on 
1992 March 31 and 400 K on 1993 March 14. The uncertainties in these estimates 
are less than 100 K, and they result mainly from uncertainties in setting
the underlying continuum level below the Balmer feature. The temperatures 
deduced are reliable for a wide range of possible density structures.  
Numerical experiments show that a uniform density sphere with radius at 
velocities ranging from 2500 to 4000 \kms\ provides satisfactory fits to both 
the 1992 March and 1993 March observations and yields practically the same 
electron temperatures for each epoch.

As an additional test of the identification of the 3630\ \AA, feature with 
the Balmer continuum profile, we have estimated the electron densities
required to produce the observed total flux in the Balmer continuum. 
Adopting the same model parameters as used in deducing the line profiles, 
the average electron densities required are found to be 
3.2$\times10^3(v/3000\,{\rm km\,s^{-1}})^{-3/2}\,{\rm cm}^{-3}$ and  
2.5$\times10^3\, (v/3000\,{\rm km\,s^{-1}})^{-3/2}\, {\rm cm}^{-3}$ for the 
1992 and 1993 data, respectively. For comparison, we can also derive electron 
densities using the \Ha\ line. Assuming that recombination
of thermal electrons is the only process responsible for the \Ha\ line, 
the electron density for both epochs is found to be 6.3$\times10^3\, 
(v/3000\,{\rm km\,s^{-1}})^{-3/2}\, {\rm cm}^{-3}$ at day 1862. This
is two to three times larger than the densities derived for the Balmer 
continuum. However, considering both the uncertainties in the absolute 
calibrations of the spectra and the over-simplified nature of the model, 
this agreement is satisfactory and adds to the evidence that our identification
is correct. The derived electron density implies that the hydrogen-rich 
region is mostly neutral with ionization fraction around 
1.2$\times10^{-4}\,({\rm M_H}/10\,{\rm M_\odot})^{-1}$.

Chugai et al. (1993) reported identification of the Paschen continuum in the 
infrared, and used the Paschen continuum to determine the temperature 
of the hydrogen emitting region. The redshifted part of the 
Paschen continuum is corrupted by some unidentified features, and the 
temperature determination is less accurate than that derived here from 
the Balmer continuum. Nonetheless, the temperatures obtained
from the Paschen continuum are consistent with what we have derived in this 
study.

\subsection{Oxygen Lines}

\subsubsection{\OID\ Doublet}

Chugai (1988) noted that in the case of supernova explosions, forbidden lines
such as \OID\ are optically thick for about one year after 
the explosion, due mainly to the high densities encountered. The doublet 
ratio $R\,=\, F([{\rm OI}]\lambda\, 6300)/F([{\rm OI}]\lambda\, 6364)$ will 
be smaller than its optically thin limit of 3. 
The ratio $R$ can be used to estimate the average density of the oxygen 
emitting region. Time evolution of the line ratio $R$ was indeed 
observed by Phillips \& Williams (1991) who found $R\,\approx\,1$ at 
about 100 days after explosion increasing to about 2.6 at about 500 
days after the explosion. Several authors have used this method to calculate 
the oxygen density in the SN 1987A ejecta (Spyromilio \& Pinto 1991; Li \& 
McCray 1992). The oxygen density derived from this method is 
$n_{OI}\, = \, 1.3\times10^9\,t_y^{-3}\,{\rm cm}^{-3}$. This is
too large if all the oxygen is distributed uniformly, and suggests that most 
of the oxygen is in dense clumps with a volume filling factor 
$f_{\rm O}\,\approx\,0.09\,M_{\rm O}/M_\odot$. Five years after the outburst, 
the density of the ejecta has greatly decreased, and the \OID\ doublet should
turn optically thin with a line ratio $R\, = \, 3$. However, the simple 
two--Gaussian 
profile fits yield a ratio $R\, = 1.9\pm 0.1$ for both the 1992 and 1993 
observations. This ratio is significantly different from the optically thin 
limit of 3, and is even smaller than the previously reported ratio of 2.6 on 
day 500 (Phillips \& Williams 1991).

If we still apply Chugai's method for density determination, we would arrive 
at 
an oxygen density of 4.1$\times 10^8\,{\rm cm}^{-3}$ on day 1862, and 
3.7$\times 10^8\,{\rm cm}^{-3}$ on day 2210. If the oxygen mass is of the
order of 1--2\Msun, and the majority of the oxygen mass is distributed inside
a shell with maximum velocity of 1700 \kms\ (as used by Li \& McCray 1992), a 
volume filling factor of $\sim\,2\times\,10^{-3}$ can be derived. This is 
about two orders of magnitude smaller than the value deduced by Li \& McCray 
(1992) using the same lines but for early observations. The latter authors 
found a volume filling factor of about 0.12 by modelling observations prior to 
day 500. Such a change could occur if the majority of the oxygen 
material is not emitting efficiently at late times, and only the 
densest parts of the ejecta are responsible for the observed line fluxes. 
In such a scenario, the early observations show the less dense part of the 
ejecta, while the later observations show the denser clumps of the ejecta 
which occupy less volume. This possibility, however, is unlikely as 
judged by the physical processes responsible for the line emission. The 
SN 1987A ejecta are expanding homologously and the density should drop as 
$t_y^{-3}$. At such late stages of the supernova ejecta, as we have derived 
above from the Balmer continuum, the electron temperature is as low as several 
hundred degrees Kelvin. This means that thermal excitations are negligible in 
powering the 
emission lines, and the \OID\ doublet photons are produced purely by 
non--thermal excitations by fast electrons from gamma ray deposition in the 
ejecta. Since the ejecta are now optically thin to gamma rays, the energy 
deposited per atomic oxygen is nearly independent of local oxygen density. 
We would then expect that all oxygen atoms are excited equally; thus, it does 
not seem likely that only the densest part in the ejecta contributes to 
the \OID\ emission. Another possibility is that the oxygen is becoming more 
clumpy with time, but this would require strong deviations from homologous
expansion of the ejecta. No physical model leads naturally to this situation. 

Blending with other lines may also be possible. Lines that are close to
[OI ] $\lambda$6364\ include those from Fe I multiplet 13, forbidden 
transitions 
of Fe I mulitplet F17, FeII multiplet 40, and Ti I multiplet 1. Line 
blending may affect the measured line ratio, but quantitative estimates 
of the effect are difficult. Here we restrict our analysis to contributions 
by a {\it single} third emission line having the same velocity structure as 
\OID. The observed line profiles are modelled by minimizing $\chi^2$  
fits of three Gaussians. We assume: (a) all three Gaussians 
have the same FWHM; (b) the contributions from \OID\ are given by two 
Gaussians with intensity ratio of 3/1 of which the weaker component is 
redshifted by 3047.6 \kms\ (64\ \AA) with respect to the blue component. 
The approximate continuum level is  simultaneously fitted by a linear function.
Figure 5a shows the results of such experiments. 
The goodness of the fit is considerably improved compared with the
two-Gaussian fit described above.
The strengths of the 
postulated third component are 14.4\% and 27.3\% of the total line fluxes in 
the 1992 and 1993 spectra, respectively. The absolute strength and the profile 
of this component did not change during the two observations. The increase 
in its relative contribution to the total line intensity was due to a 
decrease in the total line intensity. The centroid of the third Gaussian
is redshifted by 2500\,$\pm73$\,\kms\ (52\,$\pm\, 1.5$\ \AA) with respect 
to the [O I] $\lambda$6300 line. The iron line closest in wavelength is from 
Fe I $\lambda$6355.6 (multiplet 13). However, Fe I $\lambda$6355.6 should be 
blended with other Fe I multiplet 13 lines at wavelength 6402.1\ \AA, 
6360.5\ \AA, and 6282.4\ \AA.  Because the permitted Fe I lines can be optically 
thick at the time of the observations, these lines may be as strong as  
Fe I $\lambda$6355.6. The overall contribution from the Fe I multiplet 13 
should thus have a profile very different from that of a single line. The 
lines of Fe II multiplet 40, and Ti I multiplet 1, 
although having acceptable wavelengths, may have the same difficulties in 
explaining the observed line ratio. 
An alternative explanation of the abnormal line ratio is that the 
[Fe II] $\lambda\lambda$6340,6444 (15F) line is strong in SN 1987A
(Spyromilio, 1995, private comunication).  
The problem with this possibility is that the [Fe II] $\lambda$6444 
line is usually stronger than the [Fe II] $\lambda$6340, and it is too 
far to the red side of the [O I] $\lambda$6364 line to fit the observations.
Unless a reliable model of the iron emission  
process is available, it is hard to analyse the iron lines quantitively.
Nevertheless, it seems difficult to account for the observed \OID\ 
line ratio by contamination with a third component.

An alternative explanation is that the ratio deduced by line profile fitting 
is an artifact of the technique. In fact, by assuming different line 
profiles, one gets quite different values of the line ratios with   
satisfactory matches to the observed line profiles. A unique solution of the 
line ratio $R$ is never guaranteed. To be more specific, the problem 
of the doublet decomposition is described by the following 
simple equation 
$$ F(\lambda)\, =\, A_1(\lambda)\,+A_2(\lambda+\Delta\lambda_0)/R,
\eqno(3)$$
where $F$ is the flux of the \OID\ doublet, $\lambda$ is the wavelength, 
$\Delta\lambda_0$ is the wavelength difference of the two lines which 
is equal to 64\ \AA in the present case, $A_1$ and $A_2$ are the line profiles
for the 6300\ \AA\ and 6364\ \AA\ lines, respectively.  
We will assume that the \OIDB\ and \OIDR\ lines have the same line profile;
$A_1$ and $A_2$ are then identical.

Instead of using Gaussian profiles to calculate the ratio $R$, we 
reverse the process by fixing the ratio $R$ and calculating the
resulting line profile $A$. This avoids the ambiguities of unknown line 
profiles, and produce a clearer physical picture. In addition, the solution 
is mathematically unique for each value of $R$. We show in Figure 5b the 
decomposed profiles of the \OIDB\ and \OIDR\ lines together with the observed 
profile of the \MgIO\ line for comparison, assuming a line ratio $R\,=\, 3$ 
for the [O{\sc I}] doublet. It is surprising that such a simple technique 
yields \OIDB\ line profiles that are strikingly similar to those of the 
\MgIO\ line. 
This agreement suggests that the decomposed profiles are adequate
representations of each individual component of the \OID\ doublet, and argues 
strongly that the line ratio is in good agreement with the theoretical value 
expected in the optically thin limit, $R\,=\,3$. Least $\chi^2$ fit of the 
doublet ratio using the \MgIO\ line as a template yields numbers of 
3.1$\,\pm\,0.3$ for the two epochs of observations. The uncertainties 
arise largely from the difficulties in defining the continuum level for 
the lines. 

The present analysis shows that the line ratio depends sensitively on the
assumed line profile. It raises doubts concerning
the early \OID\ doublet ratios derived by line profile fitting  
(Phillips \& Williams 1991; Spyromilio \& Pinto 1992). We believe that a 
more careful analysis of the early data, considering the approaches 
outlined above, is required before drawing quantitative conclusions about 
the oxygen density in the SN 1987A ejecta. A detailed comparison of the 
early optically thick oxygen profiles with the later optically thin 
profiles is of critical importance to establish a quantitative picture of 
the oxygen emitting region in the SN 1987A ejecta.

\subsubsection{\OIIUV}

The prominent \OIIUV\ line requires some remarks. Surprisingly, this line is 
one of the strongest in the SN 1987A spectra -- in fact, second only to 
\Ha. As for the \OID\ lines, the temperature of the ejecta is too low for 
thermal electrons to excite the \OIIUV\ doublet at the epochs of these 
observations, so only non--thermal excitations of the O$^{+}$ ions are 
important. The ionization fraction of the oxygen-rich 
region five years after explosion is less than 0.001 as shown by several 
calculations (Fransson \& Chevalier 1989; Xu \& McCray 1992; Kozma \& 
Fransson 1992); therefore, only a miniscule fraction of the oxygen is ionized. 
Detailed calculations which take into account non--thermal excitations
show that the expected 
\OIIUV\ doublet strength should be less than 10 percent of that of 
the \OID\ lines (Kozma \& Fransson 1992), while the observed line strength
of \OIIUV\ is considerably larger than that of \OID. 

As discussed by Williams (1982) in his explanation of the unusual spectra 
of the shells of novae, another way to produce the 
\OIIUV\ lines may be the recombination of ${\rm O}^{++}$ ions. The 
\OIIUV\ lines are the strongest optical recombination lines from 
${\rm O}^{++}$. However, taking the effective recombination rates for 
the \OIIUV\ doublet as given by P\'{e}quignot et al. (1991), it is easy 
to verify that recombinations produce an insignificant amount of \OIIUV\  
photons.  

A more natural mechanism for powering the \OIIUV\ doublet is ionization 
of atomic oxygen directly to excited states of O$^+$
in a single impact. Updated excitation and ionization cross sections for 
electron impact on atomic oxygen by Laher \& Gilmore (1989) include  
the cross sections of collisional ionization to excited states of 
${\rm O}^+$. Their results show that for electrons with energy above 
several tens of electron volts, a significant fraction of the ionizations of 
the neutral oxygen end up with ${\rm O}^+$ ions in their $^2D^0$, $^2P^0$, 
and $^4P$ states. For example, when the energy of the free electron is 50 eV, 
about 20.6\% of the total ionized ${\rm O}^+$ ions go to the $^2D^0$ 
state, and 59.4\% go to the ground state $^4S^0$; at very high energies 
($>\,1000$ eV), the ionization to the ground state of the ${\rm O}^+$ 
line is 36\%, to $^2D^0$ is 30\%, to $^2P^0$ is 17\%, 
and to $^4P$ is 18\%. The fractions going into excited states are 
almost independent of the energy of the free electrons at energies above 
1000 eV. 

The spontaneous decay of the levels $^2D^0$, $^2P^0$, and $^4P$ of O$^+$
can be an efficient route to produce \OIIUV\ and \OIIIR\ photons. The only path
for the decay of $^2D^0$ is to the ground level with the emission of 
\OIIUV. The $^2P^0$ level decays to the ground level producing 
[OII]$\lambda$2471 or to the $^2D^0$ level with the emission of  
[OII]$\lambda$7330, with a branching ratio of 0.3775.   
For each ionization of neutral oxygen, the number of \OIIUV\ photons produced is then
$$\xi\,=\, r_{^2D^0}\,+\, 0.3775\, r_{^2P^0}, \eqno(4)$$
where $\xi$ is the fraction of ionizations to the excited states of 
${\rm O}^+$, and $r_{^2D^0}$ and $r_{^2P^0}$ are the fraction of ionizations
to the states ${^2D^0}$ and ${^2P^0}$, respectively. 
Collisional de-excitation of the $^2D^0$ level can be important,
further reducing the line flux approximately by a factor of 
$\eta\,=\,1/(1+N_e/N_{cr})$, where the critical density 
$N_{cr}\,=\,3.6\times10^{3}\, {\rm cm^{-3}}$ at a
typical temperature of 500 K (Osterbrock 1989), and $N_e$ is the 
electron density. The ionization of the 
neutral oxygen per second per cubic centimeter is 
$$ 4\pi{J_\gamma\sigma_{\gamma,OI}\over \chi_{eff,OI}}\,n_{OI},\eqno(5)$$
where $J_\gamma$ is the mean gamma ray intensity, and $\sigma_{\gamma,OI} = 
\kappa_\gamma A_{OI} m_p$ with the mass absorption coefficient 
$\kappa_\gamma=0.06 Z_{OI}/A_{OI}\, {\rm cm^{2}\,g}^{-1}$. Here $Z_{OI}=8$ and 
$A_{OI}=16$ are the atomic number and mass of the element. 

For ionization fraction $x_e\,<<\,1$, the gamma ray mean intensity is
given by (Kozma \& Fransson 1992)
$$J_\gamma\,=\,{D_\gamma L_\gamma \over 16\pi^2 (vt)^2},\eqno(6)$$
where $D_\gamma$ is a geometric factor of order unity, $v$ is the velocity 
of the ejecta, and the luminosity of the gamma rays $L_\gamma$ is given by
$$
\begin{array}{ll}
L_\gamma\,=\, & \displaystyle{
                9.1\times 10^{41} \left({M (^{56}Ni)\over 0.07 
{\rm M}_\odot}\right) \left( e^{-t/111.3}+9.3\times 10^{-4} e^{-t/391.2}\right)\,+} \cr
              &\displaystyle{
                4.1\times 10^{36} \left({M (^{44}Ti)\over 10^{-4} {\rm M}_\odot}\right)
e^{-t/28489.5} {\rm\ erg\,s^{-1}}},   
\end{array}
\eqno(7)$$
assuming 0.07 \Msun\ of $^{56}{\rm Ni}$ and $^{57}$Ni/$^{56}$Ni is 1.5 times 
the solar $^{57}$Fe/$^{56}$Fe ratio, or 2.5$\,\times\,10^{-3}$ \Msun. The 
\OIIUV\ flux is then estimated to be
$$
\begin{array}{ll}
F_{[OII]}\,= \, & \displaystyle{5.5\times 10^{-10}\,\eta\,D_\gamma\,
\left({L_\gamma \over 
9.1\times 10^{41}\, {\rm erg\,s^{-1}}}\right)\, 
              \left( {v\over 2000{\rm km\,s}^{-1}}\right)^{-2}\,\left({t\over 
              2000\,{\rm days}}\right)^{-2}}\cr
            & \displaystyle{\left({M_{\rm O}\over {\rm M}_\odot} \right)\,
              \left({D\over 50\,{\rm kpc}}\right)^{-2}\,
              {\rm erg\, cm^{-2}\,s^{-1}}},
\end{array}
\eqno(8) $$
where $D$ is the distance to the LMC, and $t$ is in units of days, and we have
taken $\chi_{eff,OI}\,=\,26.7$ eV. 

As a further test of our line identification, we
need to estimate the amount of oxygen mass required to produce the 
\OIIUV\ doublet. The line fluxes after 1500 days are also very sensitive to 
the exact amount of energy input to the ejecta, a higher energy input can 
reduce the mass of oxygen required to reproduce the observed line flux. 
Woosley \& Hoffman (1991) argue that the $^{44}$Ti mass could range from 
10$^{-6}$ to 10$^{-4} {\rm M_\odot}$. In a more recent calculation, 
Timmes et al. 
(1995) found an upper limit of $\sim$1.5$\times10^{-4}$\Msun\ of $^{44}$Ti 
that can be synthesized in a Type II supernova like SN 1987A. Late time  
observations show that although time dependent effects (Fransson \& Kozma 
1993) can reduce the amount of energy sources required to fit the bolometric
light curve, some additional energy source such as an accreting X-ray pulsar,
and/or a pulsar are still necessary to fit the observations 
(Bouchet et al. 1993).  

The observed \OIIUV\ flux corrected by interstellar extinction on day 1862 is 
8.6$\pm1.6\,\times\, 10^{-14}{\rm \,erg\,cm^{-2}\,s^{-1}}$, and on day 2210 
it is 
$7.1\pm1.4\,\times\,10^{-14}\,{\rm erg\,cm^{-2}\,s^{-1}}$.  
These numbers are reasonably reproduced by assuming a $^{44}$ Ti mass of 
4$\,\times\, 10^{-4}$ \Msun, and still keeping the mass of oxygen in a range that
is consistent with that obtained by Fransson, Houck, \& Kozma (1994).
We found from equations (7) and (8), 
$$\eta\,D_\gamma\,M_{\rm O}\,(v/2000\,km s^{-1})^{-2}\,=\,5.1\,  {\rm M_\odot} 
\eqno(9a)$$
for the 1992 observation and 
$$\eta\,D_\gamma\,M_{\rm O}\,(v/2000\,km s^{-1})^{-2}\,=\,7.6 \, {\rm M_\odot} 
\eqno(9b)$$
for the 1993 observation, respectively. 
In a simple type of analysis similar to the above, Fransson, Houck, \& 
Kozma (1994) derive a lower limit of 
$M_{\rm O}\,\ge\,3 (v/1500\, {\rm km\, s^{-1}})^{2}$ \Msun.
If we assume, as derived by Kozma 
\& Fransson (1992), that $D_\gamma\,\le\,3$, we then obtain, for typical 
velocities of the oxygen shell around 1500\ {\rm km sec$^-1$}, a lower 
limit for the oxygen mass of 
$\eta\,M_{\rm O}\,\ge\, 0.96(v/1500 \,{\rm km\,s^{-1}})^{2}$\Msun.
The correction for collisional excitation requires knowledge of the 
electron density which depends further on the clumpiness and ionization 
fraction of the ejecta. To be consistent with Fransson, Houck, \& Kozma (1993), 
we require $\eta$ to be around 0.3, 
which can be achieved if the electron density is about 
$8.4\times 10^{3}\,{\rm cm^{-3}}$. The corresponding ionization fraction is 
$2.8\times 10^{-3}f_{\rm O}$ for 3 \Msun\ of oxygen distributed
inside a sphere with maximum velocity 1500 \kms, where $f_{\rm O}$ is the 
volume filling factor of the ejecta.
It should be noted that the oxygen masses derived in the simplified approach
of Fransson, Houck, \& Kozma (1994) are larger than the 1.2--1.5\Msun derived 
from the \OID\ lines by Li \& McCray (1992) and Chugai (1994).  
In their more detailed models, a lower limit of oxygen mass of 1.5\Msun was
derived. The oxygen mass 
derived by Fransson, Houck, \& Kozma (1994) is also larger than or close to the
upper limit of the 0.24--1.6\Msun predicted by explosion models of the SN 1987A 
progenitor (Hashimoto, Nomoto, \& Shigeyama 1989; Woosley 1988; Thielemann, 
Nomoto, \& Hashimoto 1995). It remains 
to be seen if a revised analysis of both the line fluxes and profiles of the 
earlier \OID\ data will bring these numbers into agreement. However, it is 
encouraging that the order of oxygen mass required to produce the fluxes of the
\OIIUV\ line is correct even in this simple analysis. 

The \OIIUV\ line can also be used to put a limit on the electron density in 
the oxygen-rich region. One approach is to solve the ionization equilibrium time 
dependently, in a way similar to that for the \OID\ line described by 
Fransson, Houck, \& Kozma (1994). The line strengths and profiles set 
constraints on both the oxygen mass and its geometrical distribution. We 
outline here a simpler method for obtaining the electron density.
Because of the freeze-out effect, the total number of recombinations per unit
time will be larger than the total number of ionizations. An upper limit 
for the \OIIUV\ luminosity is obtained by assuming that each ionization 
to the excited states of ${\rm O^+}$ decays only via the radiative process:   
$$ \xi\,h\nu \alpha_{{\rm O}^+}\, N_{\rm OII} N_e/f_{\rm O}^2\,V\,
\ge\,L_{\rm [OII]3727},\eqno(10)$$
where $\nu$ is the frequency of the transition, $\alpha_{{\rm O}^+}$ the 
recombination coefficient of O$^+$ which 
equals 
$3.201\times 10^{-13}\,T_4^{-0.688}/(1-0.0174\,T_4^{1.707})$ with $T_4$ 
being the temperature in units of $10^4$ K (Arnaud \& Rothenflug 1985), 
and V the volume of the emitting region. Taking the observed fluxes for 
$L_{\rm [OII]3727}$, it is easy to derive from equation (10) that

$$ <N_{\rm OII}N_e>^{1/2}\, \ge\, 5400\,f_{\rm O}^{-1}\,
                        \left({v\over 2000{\rm \,km\,s^{-1}}}\right)^{-3/2}\,
                        \left({t\over 2000\,{\rm days}}\right)^{-3/2} 
                        {\rm cm^{-3}}. 
\eqno(11)$$

If we assume that mixing in the ejecta is macroscopic, equation (11) then
gives approximate values of the average electron density in oxygen-rich regions
of the ejecta. The corresponding recombination time scale of the oxygen-rich 
regions is then about 849 days, which is longer than the radioactive time 
scale, but smaller than the dynamical time scale. This confirms that time
dependent effects (Kozma \& Fransson 1993) are important at the epoch of these 
observations.

The most useful aspect of the identification of the \OIIUV\ line is that 
it provides an interesting method to estimate the amount of energy 
powering the ejecta. According to the mechanism outlined above,
the \OIIUV\ line is powered mainly by ionization followed imediately by 
excitations; because the spontaneous decay time scale of the excited 
level is only a few hours, much shorter than the radioactive time scale
or the dynamical time scale, the time-dependent effects as studied by 
Kozma \& Fransson (1993) will never be important for the late time evolution 
of the \OIIUV\ lines. The \OIIUV\ line strengths should therefore provide a 
reliable measure of the energy sources powering the late time emission
from supernova ejecta which is immune to time dependent effects. 
Here we have derived an equivalent of about $4\,\times\, 10^{-4}$ \Msun\ of 
$^{44}$Ti to power the late time spectra. The method may be complicated 
by blending
with iron lines near 3727\AA\ (Chevalier 1995, private comunications), 
in which case, it will provide an upper limit to the total amount of the 
underlying energy sources.

\subsection{Magnesium}

There are four magnesium lines observed in the spectrum. In addition to 
the semi--forbidden \MgIO\ line in the optical, the resonance lines \MgIUV\ 
and \MgIIUV\ are the strongest emission features in the ultraviolet. The 
most remarkable feature of the magnesium lines is the apparent difference 
of the line width in the ultraviolet and in the optical. Though it seems 
that all the lines show similar extended 
redshifted tails,  the overall line profiles are much broader in the 
ultraviolet than in the optical. Notably, the FWHM of the \MgIUV\ line is 
more than twice the FWHM of the \MgIO\ line, and is broader than all of the 
optical lines. Because the \MgIIUV\ lines are blended, and its red wing
blended with the \MgIUV\ line, it is impossible to deblend the lines and 
estimate their FWHM accurately. It is, however, very likely that the widths
of the \MgIIUV\ lines are comparable with that of the \MgIUV. The line 
profiles at the rest wavelength of \MgIUV\ at different epochs are shown 
in Figure 6, together with the \Ha\ lines for comparison.

The difference in line width offers a straightforward explanation for the
size difference of the SN 1987A ejecta in the optical and UV, as measured 
from the HST direct images obtained with FOC. Jakobson et al. (1994) 
showed that the SN 1987A 
ejecta are resolved by the HST. The measured size in the ultraviolet is 
twice that in the optical. Since the strongest optical lines are from decays 
of metastable levels which are optically thin, and the central wavelength 
for the UV image is at 2700\ \AA\ which is close to the resonance lines 
of magnesium in the ultraviolet, the observed differences in the emission 
line width are clearly due to the fact that the UV lines are formed at 
higher velocity further out in the ejecta. This gives a natural explanation 
of the observed size for the ejecta. 
A detailed analysis of the ultraviolet and optical images is an interesting 
problem which we will defer to a separate study.

\subsection{Iron Lines and Line Blending}

Numerous iron lines are present in the late time spectra of SN 1987A, 
although reliable identifications of these lines are difficult because the 
widths of the lines are so large that all the iron lines are strongly blended. 
However, in the ultraviolet, the Fe II emission lines from multiplet 
UV 1 and UV 2 are conspicuously strong. Iron probably makes a major
contribution to the diffuse emission complex at 
wavelengths from 3000\ \AA\ to 3850\ \AA; for instance, 
the Fe II optical multiplets 1,2,3,4,5,6,7,16, and 29 can 
be the major contributors. Fe I lines from optical multiplets 
4,5,6,20,21,23,24, and 25 may also be significant in SN 1987A; they are 
observed to be strong in the wavelength range from 3500\ \AA\ to 3900\ \AA\ in the 
narrow line quasar PHL 1092 (Bergeron \& Kunth 1980) and in the
Seyfert galaxy IZwl (Oke \& Lauer 1979). Note that the Balmer continuum and the 
[OII] $\lambda$3727\ \AA\ line lie just above this diffuse emission complex. 
The iron lines may affect the line profiles and intensities of the Balmer 
continuum and of \OIIUV. Quantitative estimates of this effect requires 
rigorous models of the iron emission which are difficult and uncertain. In 
our analysis of the Balmer continuum (\S 3.1.2) and the \OIIUV\ line 
(\S 3.2.2), we have assumed that the iron lines from 3500\ \AA\ to 
3900\ \AA\ are strongly 
blended and can be approximated by a smooth curve. This approximation is 
perhaps not too far from reality considering the fact that most of the Fe I 
and Fe II lines may be optically thick and severely blended. The emission 
feature from 4180 to 4515\ \AA\ shows three distinct peaks. Emission lines in 
the same wavelength range exist also in IZwl and are identified
with [Fe II] lines (Oke \& Lauer 1979). Some major contributors to these
lines are [Fe II] $\lambda$4244\ \AA\ (21 F), 
[Fe II] $\lambda$4287\ \AA\ (7F), 
[Fe II] $\lambda$4320\ \AA\ (21 F), and [Fe II] $\lambda$4458\ \AA\ (6F). 

The profile of \MgIUV\ may also be affected by iron lines. 
However, the HST direct images show strong evidence
that the size of the ejecta in the UV filter F275W is larger than that 
in the optical (\S 3.3; Jakobson et al. 1994). The \MgIIUV\ and \MgIUV\ lines 
contribute about 35\% to the total integrated 
light in the F275W filter. The ejecta size in the direct UV images 
can be understood with an intrinsically broader profile of the lines without 
strong iron contamination. 

\subsection{Asymmetries}

The highest detectable expansion velocities 
are given in Table 4 for the strongest lines.  
They are all measured on the redshifted side of the 
profiles. The lines that show the fastest expansion are \Ha\ and 
\MgIUV\ lines, 
with velocities up to 10,000 \kms. Such a high velocity for \MgIUV\ indicates 
clearly that a significant amount of the line flux is produced in the 
receding side of the hydrogen envelope of the ejecta. 

Comparisons between the line profiles of \Ha, \Hb, \MgIO, \OID, and \OIIUV\ 
show strong similarities. They are all blueshifted by a similar amount and 
exhibit extended redshifted wings. The blueshifts of the line peaks with 
respect to the narrow nebular lines vary from 300 to 500 \kms. Considering the 
uncertainties in the measurements and the lack of clear measurable peaks 
for many lines, these numbers are about the same as those reported earlier 
from ground-based studies which show that the peaks are blueshifted by 
about 800 and 500 \kms for \MgIO\ and [OI]6300 (Bouchet et al. 1993) on around 
day 2000.

It has been convincingly shown that the blueshift of the emission lines is
due to dust formation in the SN 1987A ejecta (Lucy et al. 1989). 
Lucy et al. (1991) show evidences of a two component structure of the
dust, diffuse small grains which produce the selective extinction at
short wavelength, and very dense dust clumps which give rise to the
wavelength-independent effect. By the time of these observations, the
ejecta had expanded several times and its density was so low that 
the diffuse component as studied by Lucy et al. (1989; 1991) were
optically thin. Applying 
the models of the line profiles of Lucy et al. (1989), we find that the 
optical depth required to account for the blueshifts in the current data is 
practically the same as in early measurements by Lucy et al. (1989; 1991). 
This property for the extinction can 
only be explained if most of the dust that condensed in the SN 1987A ejecta 
is distributed in dense clumps whose optical depth remains far above unity 
even at this late stage of evolution. Assuming that the dust clumps expand 
homologously, the optical depth then scales as $\sim t^{-2}$, and a lower 
limit for the optical depth of each individual dust clump on day 500 after 
explosion must be above 16. By requiring the dust blobs to be optically 
thick, we can derive a lower limit to the total mass of dust in the supernova 
ejecta as $$ M_{dust}\,\ge\,1.78\,\times\,10^{-5}\,{\rm M_{\odot}}\, \left({\tau\over 0.4}\right)\,
\left({\tau_b\over 1}\right)\,\left({a\over 0.1\mu {\rm m}}\right)\,\left({v\over 2000\,{\rm km\,s^{-1}}}\right)^{2}\,
\left(\rho_d\over {\rm1\, g\, cm^{-3}}\right), \eqno(12)$$
where $\tau$ is the effective optical depth of the dust blobs, $\tau_b$ is the
optical depth of each individual dust blob, $a$ is the radius of the dust 
particles, and $\rho_d$ is the density of the dust grains.
 
Inspection of Figure 5a, b and 6 showed a weak but definite 
wavelength-dependent effect of the extinction, especially for the lines 
Mg I] 4571 \AA, [O I] 6300 \AA, and \Ha. In their analysis of observations
collected at around day 550, Lucy et al. (1991) attributed this wavelength
dependence as produced by the selective extinction of the duffuse dust 
component. At around day 2000, this diffuse component should have become
optically thin and ineffective in affecting the line profiles. This may
imply that dust formation continued even at the late stage, or that the
dense dust clumps were only partially optically thick at around day 2000.

The origin of the red wing in each line profile is most likely to be the 
relic of the asymmetries observed in a much earlier phase, as shown by Phillips
\& Heathecote (1989) and Hanuschik \& Dachs (1987) for the \Ha\ line and most 
clearly by Larson et al. (1987) for P$\alpha$. Because the supernova is now 
powered mainly by the decay of $^{57}$Co and $^{44}$Ti, it thus seems that 
$^{57}$Co and $^{44}$Ti are also asymmetrically distributed, in a way quite 
similar to that of $^{56}$Co. 

\section{Conclusions}

We have analyzed the HST spectra of the ejecta of SN 1987A. The optical 
spectra of SN 1987A resemble strongly those of novae many decades after 
explosion (Williams 1992). This should not be surprising
considering the fact that the temperatures of both the SN 1987A ejecta and the 
nova shells are comparable. The hydrogen Balmer continuum yields 
accurate estimates of electron temperature in the SN 1987A ejecta. The derived 
temperatures of the hydrogen emitting region are 500K and 400K on 
day 1862 and 2210, respectively. Such low temperatures are consistent with 
model calculations of Kozma \& Fransson (1992) and Fransson \& Kozma (1994).

The \OID\ doublet is decomposed into separate [O{\sc I}] $\lambda$ 6300\ \AA\ and 
6364\ \AA\ lines assuming they have an identical profile and an intensity ratio 
of 3:1. The decomposed [O{\sc I}] profile is broadly consistent with the 
profiles of other lines such as \MgIO, \OIIUV, and 
even the hydrogen Balmer lines. The line ratios $\ROID$\  cannot be 
estimated accurately without knowing the true [O{\sc I}]$\lambda$ 6300 line 
profile. Early model calculations of the oxygen mass which used the doublet 
ratio $R$ derived on the basis of line profile fitting or line peak 
measurements are subject to this uncertainty. A robust analysis of the 
\OID\ doublet using the mechanism proposed by Chugai (1988; 1994) can only 
be obtained by data analysis which makes a careful assessment of the
line strengths without assuming a particular shape for the line profile.
This needs to be applied uniformly to both early and late data.
 
We found that the \OIIUV\ line has become surprisingly strong in the 
SN 1987A ejecta. The \OIIUV\ doublet strength is modeled by considering 
collisional ionizations of neutral oxygen directly to excited states of 
O$^+$. We note that the \OIIUV\ line potentially gives an estimate of the 
oxygen mass. Our simple analysis sets a lower limit for the SN 1987A oxygen 
mass as shown by equation (10). The mechanism responsible for the \OIIUV\ 
doublet is not unique to SN 1987A, but may be
applicable to other supernovae as well. Moreover, the mechanism is not 
unique to atomic oxygen but may also apply to other abundant elements
such as C and Ne. If this interpretation is correct, the \OIIUV\ doublet 
will to be an interesting measure of the energy sources powering late
time emission of the supernova, independent of time-dependent 
recombination. 

The \MgIUV\ and \MgIIUV\ lines in the ultraviolet form at high velocity
and large radius and thus provide a straightforward explanation for the 
size difference of the SN 1987A ejecta in the ultraviolet and in the 
optical as deduced from the HST direct images (Jakobson et al. 1992, 1994).

The ensemble of optical emission lines is blueshifted by an amount 
comparable to those reported earlier (Lucy et al. 1989, 1991), which 
requires continued high optical depth in dust from day 500 to 2000. A lower 
limit to the dust mass in the SN 1987A ejecta was derived as shown in 
equation (12). The line profiles are asymmetric; red shifted tails are 
observed for all the lines, up to 10000 \kms\ for \Ha\ and \MgIUV. The 
asymmetric tails are perhaps relics of the asymmetries observed at much 
earlier phases (Hanuschik \& Dachs 1987; Larson et al. 1987; Phillips \& 
Heathecote 1989), thus implying that $^{57}$Co and $^{44}$Ti, just 
like $^{56}$Co, are distributed asymmetrically. 

Future observations using the HST should prove to be important in studying 
the evolution of the SN 1987A ejecta. Luo, McCray, \& Slavin (1994) predict 
that the 
expanding ejecta of SN 1987A will hit the circumstellar ring in 1999$\,\pm\,3$;
a recent model by Chevalier \& Dwarkadas (1995) shows that the interaction may
start in 2005$\,\pm\,3$. When the interaction begins, the supernova will 
become a bright ultraviolet source again. The HST data provide unique 
wavelength coverage and spatial resolution and produce a large 
number of line fluxes which can be combined with models to 
extract a clearer 
picture of both the ejecta and the circumstellar ring. 

This research is supported in part by NASA Grant GO 5652, NAGW 2905, and 
NSF Grant 9218035. We are grateful for conversations with  
R. Chevalier, N. Chugai, A. M. Khokhlov, B. Leibundgut, 
J. Spyromilio, E. J. Wampler, B. Wills, and S. E. Woosley.

\clearpage
\begin{deluxetable}{ccccccc}
\footnotesize
\tablecaption{Summary of FOS Observations of SN 1987A} 
\tablewidth{0pt}
\tablehead{ 
\colhead{Dataset} & \colhead{UT Date} &\colhead{Grating} &\colhead{Detector}&\colhead{Exposure} 
&\colhead{Position}
}
\startdata
Y0WY0202T& 03/30/92& FOS/RD& G400H& 1500.0&  centered \nl
Y0WY0203T& 03/30/92& FOS/RD& G570H& 1500.0&  centered \nl
Y0WY0204T& 03/31/92& FOS/RD& G780H& 1500.0&  centered \nl
Y0WY0102T& 04/01/92& FOS/BL& G130H& 1500.0&  centered \nl
Y0WY0103T& 04/02/92& FOS/BL& G130H& 1500.0&  centered \nl
Y0WY0104T& 04/02/92& FOS/RD& G190H& 1500.0&  centered \nl
Y0WY0105T& 04/02/92& FOS/RD& G190H& 1500.0&  centered \nl
Y0WY0106T& 04/02/92& FOS/RD& G270H& 1500.0&  centered \nl
Y0WY0107T& 04/02/92& FOS/RD& G270H& 1500.0&  centered \nl
Y0WY0602T& 03/14/93& FOS/RD& G400H& 1500.0&  centered \nl
Y0WY0603T& 03/14/93& FOS/RD& G570H& 1500.0&  centered \nl
Y0WY0604T& 03/14/93& FOS/RD& G780H& 1500.0&  centered \nl
Y0WY5602T& 04/01/93& FOS/RD& G400H& 1500.0&  offcenter \nl
Y0WY5603T& 04/01/93& FOS/RD& G570H& 1500.0&  offcenter \nl
Y0WY5604T& 04/01/93& FOS/RD& G780H& 1500.0&  offcenter \nl
Y0WY0302T& 04/02/93& FOS/BL& G130H& 1500.0&  offcenter \nl
Y0WY0303T& 04/02/93& FOS/BL& G130H& 1500.0&  offcenter \nl
Y0WY0304T& 04/02/93& FOS/RD& G190H& 1500.0&  offcenter \nl
Y0WY0305T& 04/02/93& FOS/RD& G190H& 1500.0&  offcenter \nl
Y0WY0306T& 04/02/93& FOS/RD& G270H& 1500.0&  offcenter \nl
Y0WY0307T& 04/02/93& FOS/RD& G270H& 1500.0&  offcenter \nl
Y0WY0402T& 04/03/93& FOS/RD& G400H& 1500.0&  offcenter \nl
Y0WY0403T& 04/03/93& FOS/RD& G570H& 1500.0&  offcenter \nl
Y0WY0404T& 04/03/93& FOS/RD& G780H& 1500.0&  offcenter \nl
Y0WY0502T& 04/06/93& FOS/BL& G130H& 1500.0&  offcenter \nl
Y0WY0503T& 04/06/93& FOS/BL& G130H& 1500.0&  offcenter \nl
Y0WY0504T& 04/06/93& FOS/RD& G190H& 1500.0&  offcenter \nl
Y0WY0505T& 04/06/93& FOS/RD& G190H& 1500.0&  offcenter \nl
Y0WY0506T& 04/06/93& FOS/RD& G270H& 1500.0&  offcenter \nl
Y0WY0507T& 04/06/93& FOS/RD& G270H& 1500.0&  offcenter \nl
\enddata
\end{deluxetable}
\vfill
\eject

\clearpage

\begin{deluxetable}{lcccccc}
\footnotesize
\tablewidth{0pt}
\tablecaption{Observational Parameters \label{tbl-2}}
\tablewidth{0pt}
\tablehead{
\colhead{FOS Grating} & \colhead{Detector} & \colhead{Wavelength Region (\AA)} & \colhead{Dispersion\tablenotemark{a}}  & \colhead{Resolution\tablenotemark{b}}& \colhead{Aperture}
}
\startdata
G130H        & Blue     & 1140-1605               & 0.25                 & 1.1           & 0\arcsec.25$\times$2\arcsec.0\nl
G190H        & Red      & 1600-2310               & 0.36                 & 1.5           & 0\arcsec.25$\times$2\arcsec.0\nl
G270H        & Red      & 2230-3270               & 0.51                 & 2.1           & 0\arcsec.25$\times$2\arcsec.0\nl
G400H        & Red      & 3250-4791               & 0.74                 & 3.1           & 0\arcsec.25$\times$2\arcsec.0\nl
G570H        & Red      & 4200-6800               & 1.08                 & 4.1           & 0\arcsec.25$\times$2\arcsec.0\nl
G780H        & Red      & 6500-8600               & 1.43                 & 5.1           & 0\arcsec.25$\times$2\arcsec.0\nl
\enddata
\tablenotetext{a}{In units of \AA/pix}.
\tablenotetext{b}{In units of FWHM(\AA).}
\end{deluxetable}

\clearpage
\begin{deluxetable}{crrrrrrl}
\footnotesize
\tablewidth{0pt}
\tablecaption{Line List \label{tbl-3}}
\tablehead{
&\multicolumn{3}{c}{1992 Mar 30}&\multicolumn{3}{c}{1993 Mar 14}& \nl
\colhead{Wavelength(\AA)} & \colhead{Flux\tablenotemark{a}} & \colhead{Flux\tablenotemark{b}} & \colhead{FWHM(\AA)} & \colhead{Flux\tablenotemark{a}}&\colhead{Flux\tablenotemark{a}}& \colhead{FWHM(\AA)} & \colhead{Identification}
}
\startdata
 2410.3&   2.6&   6.8&  58.7&   2.0&   5.0&  60.2& Fe II (UV2) \nl
 2630.4&   3.0&   6.5&  59.6&   2.7&   5.9&  54.6& Fe II (UV1) \nl
 2828.8&  10.5&  20.6& 135.0&   6.4&  12.7& 127.3& \MgIIUV,MgIUV\nl
 3314.5&   1.8&   3.1&  50.0&   2.4&   4.1&  45.0&Fe II (1) \nl
 3476.4&   2.1&   3.5&  39.0&   2.7&   4.4&  45.1&Fe II (4,6), Fe I (6)\nl
 3620.5&  18.4&  29.7&  79.5&  15.2&  24.6&  72.1&Bac  \nl
 3737.8&  21.0&  32.8&  63.6&  24.6&  38.6&  49.7&\OIIUV\nl
 4229.3\tablenotemark{a}&      &      &      &      &      &      & [Fe II] (21F)\nl
 4335.7\tablenotemark{a}&  41.4&  55.9& 282.1&  50.4&  68.0& 284.0& H$\gamma$,[Fe II] (21F)\nl
 4456.3\tablenotemark{a}&      &      &      &      &      &      & [Fe II] (6F)\nl
 4554.3&  13.5&  17.4&  41.3&  13.9&  18.0&  39.3& \MgIO\nl
 4701.9&   1.5&   1.9&  52.9&   3.0&   3.8&  60.0& [Fe II] (5F)\nl
 4863.8&  20.7&  25.4&  86.2&  22.2&  27.3&  75.1& \Hb\nl
 5010.9&   5.1&   6.0&  45.3&   3.6&   4.3&  39.1&[Fe II] (4F)\nl
 5160.0&  11.1&  13.0&  78.6&   7.8&   9.2&  73.8&[Fe II] (18F, 35F)\nl
 5277.0&   9.9&  11.5&  93.7&   6.4&   7.3&  83.6&[Fe II] (17F, 18F, 35F)\nl
 5685.2&   9.0&   9.9& 118.2&   9.3&  10.3& 107.6&[Fe II] (33F) \nl
 5895.2&  20.1&  21.5& 118.7&  20.1&  21.6& 119.2&He I $\lambda$5876, Na I $\lambda$5896\nl
 6294.3&  17.4&  17.8&  99.9&  19.1&  19.6& 115.7&\OID\nl
 6558.8& 100.0& 100.0 &  97.1&  100.0 & 100.0 &  96.0&\Ha\nl
 7306.6&  44.7&  42.0& 125.1&  42.2&  39.6& 133.9&[O II] $\lambda$7320 , [Ca II] $\lambda$7300\ \nl
\enddata
\tablenotetext{a}{Observed flux.}
\tablenotetext{b}{Dereddened flux.}
\tablenotetext{c}{These three lines are blended.}
\end{deluxetable}

\clearpage
\begin{deluxetable}{lrr}
\footnotesize
\tablewidth{0pt}
\tablecaption{Maximum Observable Velocities for Strong Lines\label{tbl-4}}
\tablehead{
\colhead{Line}  & \colhead{1992 Mar}\tablenotemark{a}  & \colhead{1993 Mar}\tablenotemark{a} 
}
\startdata
\Ha   &  11000 & 11500                         \nl
\Hb   &  7200  & 6900                          \nl
\OIDB &  6000  & 5800                          \nl
\MgIO &  4000  & 3900                          \nl
\OIIUV&  4400  & 4750                          \nl
\MgIUV&  10000 & 10000                         \nl
Balmer Cont.   & 3100 &2700                     \nl
\enddata
\tablenotetext{a}{In units of \kms.}
\end{deluxetable}

\clearpage
\noindent
{\large\bf Figure Captions}

\noindent
{\bf Fig. 1...} The slit orientation and position for each set of 
observations.\\[3mm]
{\bf Fig. 2...} a) The 1992 March 31 spectrum and the suggested line 
identifications; b) the 1993 March 14 spectrum.
Both spectra are corrected for interstellar extinction assuming
$E(B-V)\,=\,0.2$ mag. The circumstellar lines are removed from the 
spectra.\\[3mm]
{\bf Fig. 3...} H$\alpha$ and H$\beta$ line profiles showing redshifted 
tails with velocities reaching 10000 \kms. The \Hb\ line is apparently
asymmetric but the redshifted tail is corrupted by a neighboring iron line.
Note that the narrow \Ha, \Hb, and [N II] lines are removed.
\\[3mm]
{\bf Fig. 4...} The observed hydrogen Balmer continuum profiles (solid lines)
and the corresponding model fits (dashed lines). 
\\[3mm]
{\bf Fig. 5a...} The \OID\ profiles and three-Gaussian fits. 
The parameters and constraints employed are given in the text.\\[3mm]
{\bf Fig. 5b...} The \OID\ doublet decomposed by setting the ratio 
$R\,=\,3$. The decomposed line profiles are compared with the profiles
of the \MgIO\ line for each observation.
Note that the \MgIO\ lines are arbitrarily shifted upwards
for clearity. The sum of the \OIDB\ and \OIDR\ profiles 
matches exactly the corresponding observed \OID\ profile by definition (c.f. 
equation (3)) \\[3mm]
{\bf Fig. 6...} The \MgIUV\ line is compared with the \Ha\ and \MgIO\ lines. 
From top to bottom, the profiles are \Ha, \MgIO\ and
\MgIUV. Both the \MgIUV\ and \Ha\ lines show high velocity components reaching 
10,000 \kms, but the FWHM of the \MgIUV\ line is much larger than that 
of any other observed lines.

\end{document}